\def\mi{\medskip\noindent}
\def\rf#1{(\ref{#1})}

\def\bv{{\bf b}}
\def\ia{{\bf i}_A}
\def\ib{{\bf i}_B}
\def\ic{{\bf i}_C}
\def\ifi{{\bf i}_\varphi}
\def\ir{{\bf i}_\rho}
\def\iz{{\bf i}_z}
\def\ev{{\bf e}}
\def\uv{{\bf u}}
\def\Fv{{\bf F}}
\def\Rot{\nabla\times}
\def\Div{\nabla\cdot}
\documentstyle[12pt]{article}
\begin{document}
{\center

\mi{\bf Determination of a flow generating a neutral magnetic mode}

\mi
VLADISLAV ZHELIGOVSKY

\mi
International Institute of Earthquake Prediction Theory\\
and Mathematical Geophysics\\
84/32 Profsoyuznaya St., 117997 Moscow, Russian Federation\\

\mi
Observatoire de la C\^ote d'Azur, CNRS\\
U.M.R. 6529, BP 4229, 06304 Nice Cedex 4, France\\

}

\medskip The problem of reconstruction of a flow of conducting incompressible
fluid generating a given magnetic mode is considered. We use the magnetic
induction equation to derive ordinary differential equations along the
magnetic field lines, which give an opportunity to determine the generating
flow, if additional data is provided on a two-dimensional manifold transversal
to magnetic field lines, and show that an arbitrary solenoidal vector field
can not be a neutral magnetic mode sustained by any flow of conducting fluid.

\mi
{\bf 1. Introduction}

\mi
According to the modern scientific paradigm, magnetic fields of astrophysical
objects, ranging from planets to galaxies, are often sustained by conducting
fluid flows, driven by convection in the melted medium in their interiors
\cite{Park,Moff,Zel,Pr84,RSSh}. These processes are governed
by the Navier-Stokes and magnetic
induction equations (supplemented by other equations, such as heat equation and
rheology relations, as appropriate). However, it is difficult to study them
numerically because of the extreme parameter values involved, which require
prohibitively high resolution of simulations. Thus, application of analytical
or semi-analytical methods to the study of astrophysical dynamos appears
unavoidable. In the present paper we suggest an approach, in principle enabling
one ``to separate'' the two fundamental equations; hopefully, this can be useful
for investigation of asymptotics of astrophysical dynamos.

Usually the magnetic induction equation is employed for investigation of the
evolution of magnetic field for a given flow of incompressible conducting fluid
(which is predefined in kinematic dynamo problems, or supposed to evolve
simultaneously when nonlinear dynamos are studied). We consider here an inverse
problem, investigating which consequences existence of a neutral magnetic mode
bears upon the generating flow. We show how the flow can be reconstructed
uniquely up to the data which must be provided on a two-dimensional manifold
transversal to magnetic field lines. We demonstrate that an arbitrary
solenoidal vector field can not be a magnetic mode sustained be any flow of
incompressible fluid, unless the field satisfies a consistency equation in the
fluid volume. We hope that such an analysis may be useful, in particular, for
examination of asymptotical properties of various steady magnetohydrodynamic
systems and their stability.

In recent numerical studies of nonlinear magnetic dynamos acting in plasma
\cite{DAr} and fluid \cite{Po3,CG1,CG2} flows with a prescribed forcing,
as well as in thermal convection in a horizontal layer of conducting fluid
rotating about a vertical axis \cite{Zh9b} or in the absence of rotation
\cite{Po6,Po8}, it was discovered that temporal evolution can result in
emergence of a steady state with a non-vanishing magnetic field. Magnetostatic
equilibria in ideal plasma were discussed in \cite{Bs}. A magnetic
field of a steady configuration is a neutral magnetic mode, i.e., a vector
field belonging to the kernel of the magnetic induction operator. Neutral
magnetic modes play an important r\^ole in large-scale dynamos
\cite{Zh1}-\cite{Zh9a}.

We therefore focus on neutral magnetic modes in our analysis. However,
a straightforward modification of our approach can be applied for reconstruction
of flows for eigenfunctions of the magnetic induction operator associated
with any given eigenvalue, or for arbitrary evolving magnetic fields.

\mi
{\bf 2. Reconstruction of flows}

\mi
Consider the magnetic induction equation
\begin{equation}
\partial_t\bv=\eta\nabla^2\bv+\Rot(\uv\times\bv).
\label{magind}\end{equation}
Magnetic field is solenoidal:
\begin{equation}
\Div\bv=0.
\label{bsolen}\end{equation}
In a steady state magnetic field is a neutral mode
of the magnetic induction operator. For a given flow $\bv$ and molecular
diffusivity $\eta$ the operator is elliptic. If magnetic field generation
in a bounded volume of fluid is considered and regular boundary conditions
for magnetic field are imposed, it has a discrete spectrum,
with the eigenvalues tending to $-\infty$. For a randomly chosen pair
$\eta,\uv$ the kernel of the operator does not contain mean-free magnetic
fields, and generically the only mean-free solution is $\bv=0$.

The processes bringing the system to a steady state thus can be viewed as
adjustment of the flow to a configuration allowing for a non-zero neutral
mean-free magnetic mode. It is natural therefore to treat \rf{magind}
as an equation in $\uv$. ``Uncurling'' it, one obtains
\begin{equation}
\eta\Rot\bv=\uv\times\bv-\eta\nabla a,
\label{uncurl}\end{equation}
where $a$ is a scalar function (the constant factor $\eta$ is introduced
for convenience).

Consider separately the components of \rf{uncurl} parallel and perpendicular
to $\bv$. Scalar multiplying \rf{uncurl} by $\bv$ find
\begin{equation}
(\bv\cdot\nabla)a=-\bv\cdot(\Rot\bv).
\label{aeqn}\end{equation}
The equation
controls the magnitude of a magnetic field, whose direction is prescribed:
Let $\ib$ be a unit vector collinear with $\bv$, then \rf{aeqn} implies
\begin{equation}
|\bv|=-{(\ib\cdot\nabla)a\over\ib\cdot(\Rot\ib)}.
\label{bmagn}\end{equation}
If a magnetic force line is a closed loop (including the loops emerging
due to spatial periodicity), then by virtue of \rf{aeqn}
$$\oint\ib\cdot(\Rot\bv)\,ds=-\oint(\ib\cdot\nabla)a\,ds=0$$
(the parameter $s$ on the curve is the distance along the curve from
a fixed point on it), which can be also viewed as a constraint on
the magnitude of the magnetic field (following from \rf{bmagn}\,).

The component of \rf{uncurl} perpendicular to $\bv$ is accessed by
cross-multiplication of \rf{uncurl} by $\bv$, yielding
$$\eta\,\bv\times(\Rot\bv+\nabla a)=\uv|\bv|^2-\bv(\uv\cdot\bv).$$
The component of $\uv$ parallel to $\bv$ is not determined by \rf{uncurl},
hence \hbox{$(\uv\cdot\bv)$} remains an unidentified arbitrary scalar function.
We denote
$${(\uv\cdot\bv)\over|\bv|^2}=1+\eta\alpha,$$
implying
\begin{equation}
\uv=(1+\eta\alpha)\,\bv+\eta\,\ev\times(\Rot\bv+\nabla a),
\label{vfield}\end{equation}
where
$$\ev=\bv/|\bv|^2.$$
\rf{vfield} and \rf{aeqn} together are equivalent to the equation
for a neutral magnetic mode. The scalar field $\alpha$ satisfies the equation
\begin{equation}
(\bv\cdot\nabla)\alpha+\Div\left(\ev\times(\Rot\bv+\nabla a)\right)=0,
\label{vsolenoid}\end{equation}
equivalent to the solenoidality condition for the flow $\uv$.

Now, in order to find the flow velocity \rf{vfield}, we need to determine
$\nabla a$, which we do employing the solenoidality condition for the flow.
Generically $\ib$ and $\Rot\ev$ are not parallel, and in this case
\rf{aeqn} and \rf{vsolenoid} are equivalent to the equation
\begin{equation}
\nabla a=A\ia+B\ib+C\ic,
\label{grada}\end{equation}
where $\ia,\ib,\ic$ is an orthonormal basis,
$$\ic\equiv{\Rot\ev-(\ib\cdot(\Rot\ev))\,\ib\over|\Rot\ev-(\ib\cdot(\Rot\ev))\,\ib|},
\qquad\ia\equiv\ib\times\ic,$$
\begin{equation}
B\equiv-\ib\cdot(\Rot\bv)\equiv-|\bv|\,\ib\cdot(\Rot\ib),
\label{Bdef}\end{equation}
\begin{equation}
C={(B/|\bv|)^2-(\bv\cdot\nabla)\alpha-\Div(\ev\times(\Rot\bv))
\over|\Rot\ev-(\ib\cdot(\Rot\ev))\,\ib|}.
\label{alfa}\end{equation}
We derive from \rf{grada} individual equations in $A$ and $C$.

The solvability condition for \rf{grada} is obtained by taking its curl:
\begin{equation}
0=\nabla A\times\ia+A\Rot\ia+\nabla B\times\ib+B\Rot\ib+
\nabla C\times\ic+C\Rot\ic.
\label{rotgrad}\end{equation}
Scalar multiplying it by $\ia,\ib$ and $\ic$, one finds
\begin{equation}
A={-C\ia\cdot(\Rot\ic)-(\ib\cdot\nabla)C
-B\ia\cdot(\Rot\ib)+(\ic\cdot\nabla)B\over\ia\cdot(\Rot\ia)},
\label{A}\end{equation}
\begin{equation}
0=A\ib\cdot(\Rot\ia)+(\ic\cdot\nabla)A-{B^2\over|b|}
+C\ib\cdot(\Rot\ic)-(\ia\cdot\nabla)C,
\label{B}\end{equation}
\begin{equation}
C={-A\ic\cdot(\Rot\ia)+(\ib\cdot\nabla)A
-B\ic\cdot(\Rot\ib)-(\ia\cdot\nabla)B\over\ic\cdot(\Rot\ic)},
\label{C}\end{equation}
where $B$ is defined by \rf{Bdef}.
Substitution of \rf{A} into \rf{C} yields a second order differential equation
along magnetic force lines, in principle, defining $C$. Initial conditions
for this equations must be set on two-dimensional manifolds, transversal
to magnetic force lines. They must assure geometric consistency: the solutions
along closed force lines must be periodic. For force lines, intersecting with
the boundary of the region occupied by the fluid, it is naturally to set the
conditions on the boundary. The data can be provided on two manifolds, crossing
a force line; in this case one obtains a boundary value problem for $C$.
In turn, $\alpha$ can be found, in principle, from \rf{alfa}. This completes
reconstruction of the flow. (The divergence of \rf{grada} yields an equation
in $a$, which can be used to find the potential itself.) Substituting $A$ \rf{A} and
$C$ into \rf{B}, one obtains an equation in $\bv$. Thus, not every solenoidal
field can be a magnetic neutral mode: The scalar consistency equation \rf{B}
constrains, together with the solenoidality condition, a neutral mode up to
a scalar field.

Implementation of this program can become particularly difficult in the
presence of magnetic nulls, i.e., points, where magnetic field vanishes.
(This is clear, of course, already from the definition of the vector field
$\ev$, which becomes singular at the nulls). Topology of magnetic field
with null points and its bifurcations during reconnections are studied
in detail in solar magnetohydrodynamics \cite{LF90,Lau93,Bs00,PF00,Rec} --
they are presumed to be of fundamental importance for occurrence of sudden
explosive energy release events, solar flares, in the Sun's corona.
In the vicinity of a null point magnetic field exhibits an approximately
linear behavior controlled by the Jacobian $\|\partial b_i/\partial x_j\|$.
Solenoidality of the magnetic field implies, that the sum of the three
eigenvalues of this matrix vanishes. Hence, generically it has two eigenvalues
with real parts of the same sign, and an eigenvalue of an opposite sign.
Consequently, one can identify a two-dimensional manifold of magnetic force
lines behaving coherently -- all approaching the null point or all departing from
it (if the two eigenvalues have negative or positive real parts, respectively)
and an one-dimensional manifold (a force line), exhibiting the behavior of the
opposite kind. In the parlance of solar physics, the two-dimensional manifold
is the {\it fan}, and the one-dimensional manifold the {\it spine} of
the null (see Fig.~1 in \cite{BrPr}). Therefore, in our problem
there are infinitely many characteristics (constituting the fan), which must
bring the same values of $A$ and $C$ to (or take the same values from)
the null point, implying that the problem of consistency of the global
solution for the flow arises. The situation is further complicated by
the fact that $\ib$ is typically discontinuous at null points (its direction
is not well-defined), and hence $\ia$ and $\ic$ are discontinuous as well.

Thus, the presence of magnetic null points is likely to result in a
discontinuity of the reconstructed flow, but they are not the only source
of troubles. More generally, our formalism becomes ill-defined at
the points, where the magnetic field $\bv$ is parallel to $\Rot\ev$.
If a magnetic force line crosses the boundary at two points, a problem arises
in satisfying the boundary conditions for the flow at the two points.

\mi
{\bf 3. Axisymmetric magnetic neutral modes}

\mi
Equations \rf{A}-\rf{C} suggest that the complexity of the problem depends
considerably on the geometry of magnetic force lines. For instance,
reconstruction of the flow is difficult, if
force lines exhibit a chaotic spatial behavior. We consider here one of
the simplest examples of an axisymmetric magnetic neutral mode
$$\bv=b(\rho,z)\ifi,\qquad\ib=\ifi,$$
$(\rho,\varphi,z)$ being a cylindrical coordinate system and
$\ir,\ifi,\iz$ the respective unit vectors.

Before we formulate the system of equations \rf{A}-\rf{C} in the variables $A$
and $C$, which we need to solve in order to reconstruct the flow \rf{vfield},
we derive some useful properties of the basis $\ia$, $\ib$, $\ic$.
Curls of azimuthal and poloidal vector fields independent of $\varphi$ are,
respectively, poloidal and azimuthal; this implies the orthogonality
\begin{equation}
\ia\cdot(\Rot\ia)=\ib\cdot(\Rot\ib)=\ib\cdot(\Rot\ev)=\ic\cdot(\Rot\ic)=0.
\label{orto}\end{equation}
By a simple calculation,
$$\ic={\Rot\ev\over|\Rot\ev|}=h\left(-{\partial\kappa\over\partial z}\ir
+{\partial\kappa\over\partial\rho}\iz\right),$$
where
$$\kappa(\rho,z)\equiv{\rho\over b},\qquad h(\rho,z)\equiv{1\over|\nabla\kappa|};$$
hence
$$\ia\equiv\ib\times\ic=h\nabla\kappa.$$
Therefore,
$$\ic\cdot(\Rot\ia)=\ic\cdot(\nabla h\times\nabla\kappa)=0,$$
since none of the factors in the triple product has an azimuthal component.
By vector algebra identities,
$$\ia\cdot(\Rot\ic)-\ic\cdot(\Rot\ia)=-\Div(\ia\times\ic)=\Div\ifi=0,$$
implying
\begin{equation}
\ia\cdot(\Rot\ic)=\ic\cdot(\Rot\ia)=0.
\label{ortt}\end{equation}

Now, scalar multiplying \rf{rotgrad} by $\ia,\ib$ and $\ic$ and employing \rf{orto}
(in particular, $B=0$) and \rf{ortt}, one obtains equations
\begin{equation}
0={\partial C\over\partial\varphi},
\label{Aaxi}\end{equation}
\begin{equation}
0=A\ib\cdot(\Rot\ia)+(\ic\cdot\nabla)A
+C\ib\cdot(\Rot\ic)-(\ia\cdot\nabla)C,
\label{Baxi}\end{equation}
\begin{equation}
0={\partial A\over\partial\varphi}
\label{Caxi}\end{equation}
(which are now significantly simpler than \rf{A}-\rf{C} in the general case).
Equations \rf{Aaxi} and \rf{Caxi} are equivalent to
\begin{equation}
C=C(\rho,z),
\label{Crz}\end{equation}
\begin{equation}
A=A(\rho,z).
\label{Arz}\end{equation}

For an axisymmetric magnetic field, \rf{alfa} takes the form
$${\partial\alpha\over\partial\varphi}=-|\Rot\ev|C-\Div(\ev\times(\Rot\bv)).$$
Consequently, \rf{Crz} and geometric consistency ($2\pi$-periodicity
of $\alpha$ in $\varphi$) imply that\break $\alpha=\alpha(\rho,z)$
is an arbitrary function (together with the relations \rf{Crz}, \rf{Arz} and
$B=0$, this formally confirms a physically obvious fact, that a flow
generating an axisymmetric magnetic field is necessarily axisymmetric), and
\rf{Crz} is superceded by
\begin{equation}
C=-{\Div(\ev\times(\Rot\bv))\over|\Rot\ev|}.
\label{Cfinal}\end{equation}

Now $A$ must be determined from \rf{Baxi}. We introduce characteristics
$(R(s),Z(s))$ in the $(\rho,z)$ half-plane; they satisfy the ODE's
$${dR\over ds}=-h(R(s),Z(s)){\partial\kappa\over\partial z}(R(s),Z(s)),$$
$${dZ\over ds}=h(R(s),Z(s)){\partial\kappa\over\partial\rho}(R(s),Z(s)).$$
Direct differentiation shows that the characteristics are isolines of
the scalar field $b(\rho,z)$. Since along a characteristic
$$\ib\cdot(\Rot\ia)=
{\partial h\over\partial z}{\partial\kappa\over\partial\rho}-
{\partial h\over\partial\rho}{\partial\kappa\over\partial z}=
{1\over h}\left({\partial h\over\partial z}{dZ\over ds}+
{\partial h\over\partial\rho}{dR\over ds}\right)={1\over h}{dh\over ds},$$
\rf{Baxi} takes the form
$${d\over ds}(Ah)=f,$$
where
$$f\equiv h((\ia\cdot\nabla)C-C\ib\cdot(\Rot\ic))$$
and $C$ is given by \rf{Cfinal}. Consequently,
\begin{equation}
A(R(s),Z(s))={A(R(0),Z(0))h(R(0),Z(0))
+\int_0^sf(R(s'),Z(s'))\,ds'\over h(R(s),Z(s))}.
\label{solchar}\end{equation}
If a characteristic is a closed orbit of period $S$, geometric consistency
implies that over this orbit
\begin{equation}
\int_0^Sf(R(s'),Z(s'))\,ds'=0.
\label{period}\end{equation}
Thus, we have determined $\nabla a$ and the flow \rf{vfield} (to the extent
this is permitted by the natural non-uniqueness of solutions to \rf{magind}
in $\uv$).

The well-known Cowling antidynamo theorem states that generation of smooth
axisymmetric magnetic fields (including steady ones) of finite total energy is
impossible. Two proofs of the theorem (following \cite{Cow} and \cite{Bra})
are presented in \cite{Moff}. The demonstrations rely on the equation of total
magnetic energy balance derived for a smooth axisymmetric flow
of incompressible fluid, provided the normal component of velocity vanishes on
the boundary of the region where the fluid resides. To reconcile our results
with the Cowling theorem, we note that the flow that we obtain will not satisfy
some of these conditions. It may be singular on the circles, where $b=0$, or
$\kappa$ has extrema (and then $\ev$ or $h$ are singular, respectively).
If the volume occupied by the flow is bounded, it cannot be guaranteed that
the normal component of the fluid velocity vanishes everywhere on the boundary
(or, alternatively, enforcing this condition creates a discontinuity in the
flow). Hence, the standard procedure employed to establish the total magnetic
energy balance equation will reveal additional sources of magnetic energy,
which emerge because the flow is not smooth or the surface integral
representing the contribution of the advective term does not vanish; under
such circumstances the Cowling theorem is unapplicable.

We have presented the analysis of this section mainly as an illustration
of how the proposed formalism might be applied to reconstruct flows for less
trivial magnetic field configurations. However, in addition, it provides useful
information in regard to the following technical issue: Although we have stated
at the end of the previous section that \rf{B} is a constraint for a neutral
magnetic mode, we have not yet produced any evidence, that the three equations
\rf{A}-\rf{C} are independent. Eqns.~\rf{Aaxi}-\rf{Caxi}, which we have derived
considering this particular example, demonstrate that \rf{B} is not
a consequence of \rf{A}, \rf{C} and solenoidality of magnetic field.

\pagebreak
\mi
{\bf 4. Concluding remarks}

\mi
We have shown in Section 2 that reconstruction of an incompressible flow
\rf{vfield} from the structure of a magnetic mode consists of solution of
equations \rf{A} and \rf{C} in $A$ and $C$, followed by solution of \rf{alfa}
in $\alpha$. These equations are ordinary differential equations along magnetic
force lines; thus, the problem becomes complex, if the force lines exhibit
a chaotic behavior. For a solenoidal vector field to be a neutral magnetic
mode, it must satisfy the constraint \rf{B}.

Substitution of \rf{vfield} into the momentum equation
$$\nu\nabla^2\uv+\uv\times(\Rot\uv)-\bv\times(\Rot\bv)-\nabla p+\Fv=0$$
yields an equation in $\bv$:
$$\nu\nabla^2((1+\eta\alpha)\,\bv+\eta\ev\times(\Rot\bv+\nabla a))$$
$$+\,\eta((1+\eta\alpha)\,\bv+\eta\ev\times(\Rot\bv+\nabla a))
\times(\Rot(\alpha\bv+\ev\times(\Rot\bv+\nabla a)))$$
\begin{equation}
+\,\eta(\alpha\bv+\ev\times(\Rot\bv+\nabla a))\times(\Rot\bv)
-\nabla p+\Fv=0\phantom{\int^|_|}
\label{mom}\end{equation}
comprising a closed system of equations together with the solenoidality
condition \rf{bsolen}. Relation \rf{B} now becomes a constraint on the
acceptable fluid forcing $\Fv$.

Analysis of the dependence of steady or evolving magnetohydrodynamic systems
on small viscosity and magnetic diffusivity is a notoriously difficult problem.
The structure of \rf{mom} may turn out to be advantageous for the study of
asymptotics of MHD steady states, when the force $\Fv$ is of the order of
small quantities $\nu\sim\eta$, as it is in nonlinear dynamos with energy
equipartition \cite{CG1,CG2}. (The form of the scalar factor in front of $\bv$
in \rf{vfield} has been chosen so that all terms in \rf{mom} were in this case
of the same order of smallness.)

In Section 3 we have considered an example of the reconstruction problem for
axisymmetric neutral magnetic modes. This particular case has proved to be
highly degenerate: the denominators in \rf{A} and \rf{C} vanish identically,
and the respective components of \rf{rotgrad} just testify that $\nabla a$ is
an axisymmetric vector field. Relation \rf{B} does not constrain further
the structure of the magnetic field, but rather defines, by \rf{solchar},
the component $A$ of $\nabla a$. Initial conditions $A(R(0),Z(0))$ for solutions
\rf{solchar} of \rf{Baxi} along characteristics can be chosen on curves
in the $(\rho,z)$ half-plane, which are transversal to magnetic force lines.
The azimuthal component of the flow velocity, $(1+\eta\alpha)b$, is an
arbitrary axisymmetric scalar field (in this case it is controlled neither
by the magnetic induction equation, nor, due to independence of $\varphi$,
by the solenoidality condition). Thus the reconstructed flow is unique up
to the data which must be specified on two-dimensional manifold(s) (the scalar
field $\alpha$ on the $(\rho,z)$ half-plane) and on one-dimensional
curve(s) on this half-plane (the initial conditions $A(R(0),Z(0))$\,).

The initial data must be smooth so that the resultant field $A$ had no
singularities. If the topology of isolines of the magnitude of magnetic field
$b$ is non-trivial, the smoothness of the initial data is insufficient;
for instance, geometric consistency requires that the integral \rf{period} over
any closed magnetic force line vanishes. If the axis of symmetry intersects
with the volume occupied by the fluid, axisymmetry gives rise to another
problem: regularity of the magnetic field implies $b(0,z)=0$; consequently,
the term $\ev\times(\Rot\bv)$ in \rf{vfield} tends to infinity for $\rho\to0$.
Thus, the flow is non-singular only, if initial conditions for $A$ compensate
for this singularity.

\mi{\bf Acknowledgments}

Part of this research was carried out during my visit to the School of
Engineering, Computer Science and Mathematics, University of Exeter, UK,
in January -- April 2008. I am grateful to the Royal Society for their financial
support. My research visits to Observatoire de la C\^ote d'Azur were supported
by the French Ministry of Education. My research was partially financed
by the grants ANR-07-BLAN-0235 OTARIE from Agence nationale de la recherche,
France, and 07-01-92217-CNRSL{\Large\_}a from the Russian foundation for basic
research. I am grateful to Andrew Gilbert for discussions.

\mi
\end{document}